\documentclass[twocolumn,twoside,lettersize]{IEEEtran}

\usepackage{setspace,cite}

\usepackage{graphicx}
\usepackage{multirow,multicol}

\usepackage[tbtags]{amsmath}
\usepackage{amsbsy}
\usepackage{amssymb}
\usepackage{amsfonts}
\usepackage{amsthm}

\usepackage{balance}
\usepackage{fancyvrb}
\usepackage{verbatim}

\newtheorem{theorem}{Theorem}
\newtheorem{proposition}{Proposition}
\newtheorem{lemma}{Lemma}

\newtheorem{assumption}{Assumption}

{\begin{list}               
    {$\bullet$ \hfill}{
        \setlength{\leftmargin}{\parindent}
        \setlength{\parsep}{0.04\baselineskip}
        \setlength{\itemsep}{0.5\parsep}
        \setlength{\labelwidth}{\leftmargin}
        \setlength{\labelsep}{0em}}
    }
{\end{list}}

\providecommand{\eref}[1]{\eqref{#1}}  
\providecommand{\cref}[1]{Chapter~\ref{#1}}

\providecommand{\fref}[1]{Figure~\ref{#1}}

\providecommand{\R}{\ensuremath{\mathbb{R}}}

\providecommand{\E}{\ensuremath{\mathbb{E}}}

\providecommand{\bydef}{\overset{\text{def}}{=}}

\renewcommand{\vec}[1]{\ensuremath{\boldsymbol{#1}}}
\providecommand{\mat}[1]{\ensuremath{\boldsymbol{#1}}}

\providecommand{\calD}{\mathcal{D}}

\providecommand{\calL}{\mathcal{L}}

\providecommand{\calN}{\mathcal{N}}
\providecommand{\calO}{\mathcal{O}}

\providecommand{\mA}{\mat{A}}

\providecommand{\mD}{\mat{D}}

\providecommand{\mG}{\mat{G}}

\providecommand{\mI}{\mat{I}}

\providecommand{\mK}{\mat{K}}

\providecommand{\mP}{\mat{P}}
\providecommand{\mQ}{\mat{Q}}

\providecommand{\mS}{\mat{S}}

\providecommand{\mU}{\mat{U}}

\providecommand{\mW}{\mat{W}}

\providecommand{\vb}{\vec{b}}

\providecommand{\vu}{\vec{u}}
\providecommand{\vv}{\vec{v}}

\providecommand{\vx}{\vec{x}}
\providecommand{\vy}{\vec{y}}
\providecommand{\vz}{\vec{z}}

\providecommand{\mLambda}{\mat{\Lambda}}

\providecommand{\mSigma}{\mat{\Sigma}}

\providecommand{\vepsilon}{\vec{\epsilon}}

\providecommand{\veta}{\vec{\eta}}

\providecommand{\vmu}{\vec{\mu}}

\providecommand{\mDtilde}{\mat{\widetilde{D}}}

\providecommand{\vvtilde}{\boldsymbol{\widetilde{v}}}
\providecommand{\vxtilde}{\boldsymbol{\widetilde{x}}}

\providecommand{\vxhat}{\boldsymbol{\widehat{x}}}
\providecommand{\vuhat}{\boldsymbol{\widehat{u}}}
\providecommand{\vvhat}{\boldsymbol{\widehat{v}}}
\providecommand{\vubar}{\boldsymbol{\bar{u}}}

\providecommand{\vzero}{\vec{0}}
\providecommand{\vone}{\vec{1}}

\newcommand{\subjectto}{\mathop{\mbox{subject\, to}}}
\newcommand{\argmin}[1]{\mathop{\underset{#1}{\mbox{argmin}}}}

\newcommand{\minimize}[1]{\mathop{\underset{#1}{\mathrm{minimize}}}}

\graphicspath{{pix/}}
\usepackage[pagebackref=true,breaklinks=true,letterpaper=true,colorlinks,bookmarks=false]{hyperref}

\begin{document}
\title{Performance Analysis of Plug-and-Play ADMM: \\ A Graph Signal Processing Perspective}

\author{Stanley~H.~Chan,~\IEEEmembership{Senior Member,~IEEE}
        \thanks{The author is with the School of Electrical and Computer Engineering, and the Department of Statistics, Purdue University, West Lafayette, IN 47907, USA. Email: \texttt{ stanchan@purdue.edu}.}
    }

\maketitle

\begin{abstract}
The Plug-and-Play (PnP) ADMM algorithm is a powerful image restoration framework that allows advanced image denoising priors to be integrated into physical forward models to generate high quality image restoration results. However, despite the enormous number of applications and several theoretical studies trying to prove convergence by leveraging tools in convex analysis, very little is known about why the algorithm is doing so well. The goal of this paper is to fill the gap by discussing the performance of PnP ADMM. By restricting the denoisers to the class of graph filters under a linearity assumption, or more specifically the symmetric smoothing filters, we offer three contributions: (1) We show conditions under which an equivalent maximum-a-posteriori (MAP) optimization exists, (2) we present a geometric interpretation and show that the performance gain is due to an intrinsic pre-denoising characteristic of the PnP prior, (3) we introduce a new analysis technique via the concept of consensus equilibrium, and provide interpretations to problems involving multiple priors.
\end{abstract}


\IEEEpeerreviewmaketitle

\section{Introduction}
The Plug-and-Play (PnP) ADMM is a variant of the alternating direction method of multiplier (ADMM) algorithm. Since its introduction in 2013 \cite{Venkatakrishnan_Bouman_Wohlberg_2013}, the algorithm has demonstrated extremely promising results in image restoration and signal recovery problems \cite{Dar_Bruckstein_Elad_2016, Metzler_Maleki_Baraniuk_2016,Rond_Giryes_Elad_2016,Kamilov_Mansour_Wohlberg_2017,Chang_Marchesini_2016,Ono_2017,Sreehari_Venkatakrishnan_Wohlberg_2016}. However, despite the enormous number of applications and several studies on its convergence \cite{Sreehari_Venkatakrishnan_Wohlberg_2016, Chan_Wang_Elgendy_2017, Teodoro_Bioucas_Figueiredo_2019}, it is generally unclear why the algorithm is performing so well. On the one hand, one can argue that the superior performance is attributed to the underlying image denoisers, which is evident from applications using convolutional neural networks \cite{Heide_Steinberger_Tsai_2014,Zhang2017_cvpr,Zhang2017_tip,Borgerding_Schniter_Rangan_2017}. Yet on the other hand, unless we can explicitly write down the optimization, e.g., in the form of maximum-a-posteriori (MAP), it will be extremely difficult to quantify the solution even if we know the solution exists. The goal of this paper is to present a set of analytic results so that we can understand the behavior of PnP ADMM. To do so, we consider a subset of image denoisers known as the graph filters or symmetric smoothing filters. Under such setting, we are able to explicitly show the geometry of the algorithm and make comparison with graph Laplacian based restoration algorithms.

In order to state the problem more concretely, it would be useful to first review the PnP ADMM. The starting point of the PnP ADMM is the ADMM algorithm, which has become a standard tool for minimizing a sum of two separable functions of the form
\begin{equation}
\vxhat = \argmin{\vx} \;\; f(\vx) + \lambda g(\vx).
\label{eq:basic optimization}
\end{equation}
Here, $f$ represents the \emph{objective function} arising from the physical measurement models, e.g., blur, sampling, projection, or other linear transformations. The function $g$ is the \emph{regularization function}, representing the prior model of the underlying signals, e.g., $\ell_p$-norm, sparsity, or total variation. The parameter $\lambda$ is the regularization constant, and is assumed known a-priori and is fixed in this paper. Problems taking the form of \eref{eq:basic optimization} is broad, encompassing various linear inverse problems in imaging, deblurring, super-resolution, CT, MRI, and compressed sensing, to name a few.

The ADMM algorithm solves \eref{eq:basic optimization} by converting the unconstrained optimization into a constrained problem
\begin{equation}
(\vxhat,\vvhat) = \argmin{\vx,\vv} \;\; f(\vx) + \lambda g(\vv), \;\; \subjectto \;\; \vx = \vv.
\label{eq:basic optimization 2}
\end{equation}
It then considers the augmented Lagrangian function:
\begin{equation}
\calL(\vx,\vv,\vu) = f(\vx) + \lambda g(\vv) + \vu^T(\vx-\vv) + \frac{\rho}{2}\|\vx - \vv\|^2.
\label{eq:augmented lagrangian function}
\end{equation}
Then, the algorithm finds the solution by seeking a saddle point of $\calL$, which involves solving a sequence of subproblems in the form
\begin{align}
\vx^{(k+1)} &= \argmin{\vx\in \R^n} \;\; f(\vx) +\frac{\rho}{2} \|\vx - \vxtilde^{(k)}\|^2, \label{eq:ADMM2,x}\\
\vv^{(k+1)} &= \argmin{\vv\in \R^n} \;\; \lambda g(\vv) + \frac{\rho}{2}\|\vv - \vvtilde^{(k)}\|^2,\label{eq:ADMM2,v}\\
\vubar^{(k+1)} &= \vubar^{(k)} + (\vx^{(k+1)} - \vv^{(k+1)}),\label{eq:ADMM2,u}
\end{align}
where $\vubar^{(k)} \bydef (1/\rho)\vu^{(k)}$ is the scaled Lagrange multiplier, $\vxtilde^{(k)} \bydef \vv^{(k)}-\vubar^{(k)}$ and $\vvtilde^{(k)} \bydef \vx^{(k+1)}+\vubar^{(k)}$ are the intermediate variables. Under mild conditions, e.g., when both $f$ and $g$ are closed, proper and convex, and if a saddle point of $\calL$ exists, one can show that the iterates returned by \eref{eq:ADMM2,x}-\eref{eq:ADMM2,u} converges to the solution of \eref{eq:basic optimization}. Readers interested in knowing more about the theoretical properties of ADMM can consult tutorials such as \cite{Boyd_Parikh_Chu_Peleato_Eckstein_2011}.

The idea of PnP ADMM is to modify \eref{eq:ADMM2,v} by observing that it is a denoising step if we treat $\vvtilde^{(k)}$ as a ``noisy'' version of $\vv$ and $g(\vv)$ as a regularization for $\vv$. Based on this, we can replace \eref{eq:ADMM2,v} by a denoiser $\calD_\sigma: \R^n \rightarrow \R^n$ such that
\begin{equation}
\vv^{(k+1)} = \calD_\sigma\left( \vvtilde^{(k)} \right),
\label{eq:ADMM2,v rewrite 2}
\end{equation}
where $\sigma = \sqrt{\lambda/\rho}$ is the denoising strength or a hypothesized ``noise level''. The choice of $\calD_\sigma$ is broad. $\calD_\sigma$ can be a proximal map such as total variation denoising, or an off-the-shelf image denoisers such as BM3D, non-local means, and more recently neural networks.

Replacing the explicit optimization in \eref{eq:ADMM2,v} by an off-the-shelf denoiser in \eref{eq:ADMM2,v rewrite 2} leads to many interesting questions:
\begin{itemize}
\item \textbf{Existence of $g$}. Given an arbitrary denoiser $\calD_\sigma$, is it possible to have an equivalent $g$ such that \eref{eq:ADMM2,v} and \eref{eq:ADMM2,v rewrite 2} coincide? If not, under what conditions of $\calD_\sigma$ would this happen? Some recent studies provide an answer, e.g., \cite{Sreehari_Venkatakrishnan_Wohlberg_2016, Reehorst_Schniter_2018, Teodoro_Bioucas_Figueiredo_2019}, which will be elaborated in later sections of this paper. An alternative approach is the Regularization by Denoising (RED) \cite{Romano_Elad_Milanfar_2017}, which has advantages and limitations. Other variations include \cite{Teodoro_Bioucas_Figueiredo_2019,Bigdeli_Zwicker_Favaro_2017, Tirer_Giryes_2019}.
\item \textbf{Convergence}. Does PnP ADMM converge? Majority of the known results are based on a classical result of Moreau \cite{Moreau_1965}, which suggests that global convergence of PnP ADMM is guaranteed if and only if $\calD_{\sigma}$ has symmetric gradient and is non-expansive (See \cite{Sreehari_Venkatakrishnan_Wohlberg_2016} and \cite{Teodoro_Bioucas_Figueiredo_2019}). However, non-expansiveness and symmetric gradient are strong conditions that do not hold for all denoisers. A relaxed criteria by fixed point convergence allows one to prove convergence for a slightly broader class of denoisers which are asymptotically invariant as $\sigma \rightarrow 0$ \cite{Chan_Wang_Elgendy_2017}. Fixed point convergence does not guarantee converging to a solution of certain optimization (because such optimization may not even exist), yet practically the fixed point can correspond to a meaningful image when the parameters are set properly.
\item \textbf{Performance}. If $g$ does exist, what is it and why is it an effective regularization function? An earlier work of Chan showed the form of $g$ for symmetric smoothing filters \cite{Chan_2016}, and a similar result is recently reported by \cite{Teodoro_Bioucas_Figueiredo_2019}. We will elaborate more on this later. If $g$ does not exists, it would be important to know what does PnP ADMM solve, and why it is good. The focus of this paper is to address this performance issue. We will take two paths, one through the MAP formulation, and the other one through a concept called the consensus equilibrium.
\item \textbf{Generalization}. How to generalize the algorithm to multiple denoisers? The generalization can be done by a technique called consensus equilibrium \cite{Buzzard_Chan_Bouman_2017}. However, can we explain the performance, and how to determine the optimal combination weights?
\end{itemize}
To enable our study, we assume that the denoisers are in the family of symmetric smoothing filters \cite{Milanfar_2013a, Milanfar_2013b, Chan_Zickler_Lu_2017} or graph filters \cite{Meyer_Shen_2012,Talebi_Zhu_Milanfar_2013,Taylor_Meyer_2012,Kheradmand_Milanfar_2014,Talebi_Milanfar_2014,Pang_Cheung_2017,Bai_Cheung_Liu_2018}. These filters are simple in the sense that they are (pseudo) linear and hence they can be expressed using matrices. However, they are also representative enough to cover a sufficiently large group of image denoisers that we encounter in the literature.

The remaining of the paper is arranged as follows. We begin with a brief review of the construction of graph filters and their properties in Section 2. The main results are presented in Section 3 and 4. Section 3 analyzes the problem from a MAP perspective, whereas Section 4 approaches the problem from the consensus equilibrium perspective. Further discussions of open problems are presented in Section 5.

\section{Graph Filters}
\subsection{Constructing a Graph Filter}
Graph filter is a class of linear denoisers originally developed in the area of graph signal processing. The application of graph filters is very broad, e.g., deblurring using graph Laplacian \cite{Kheradmand_Milanfar_2014}, boosting image denoisers \cite{Talebi_Zhu_Milanfar_2013}, image denoising using global similarity \cite{Talebi_Milanfar_2014}, JPEG compression using random walk \cite{Liu_Cheung_Wu_2017}, and  blind deconvolution using reweighted graph total variation \cite{Bai_Cheung_Liu_2018}. In terms of theory, there are studies of the graphs in continuous domain \cite{Pang_Cheung_2017}, sampling of graphs \cite{Anis_Gadde_Ortega_2016}, and filter banks of graphs \cite{Tremblay_Goncalves_Borgnat_2017}, to name a few.

Represented as matrices, graph filters take the form of
\begin{equation}
\vxhat = \calD_{\sigma}(\vy) = \mW \vy.
\label{eq: filtering output}
\end{equation}
We call the matrix $\mW \in \R^{n \times n}$ a graph filter. There are multiple ways of constructing $\mW$. One of the most commonly used approaches is to define a non-local weight kernel $\mK \in \R^{n \times n}$:
\begin{equation}
[\mK]_{ij} = \exp\left\{-\frac{\|\vy_i - \vy_j\|^2}{2h^2}\right\},
\label{eq: K}
\end{equation}
where $\vy_i \in \R^d$ is a $d$-dimensional patch centered at pixel $i$, and $h$ is a parameter characterizing the denoising strength of the filter. Modifications to \eref{eq: K} are common, e.g., by incorporating spatial distance (which gives the non-local means filter), restricting to pixels instead of patches (which gives the bilateral filter), or extending the norm to a weighted norm (which gives the kernel regression filter). For more examples of the kernel matrices, we refer the readers to \cite{Milanfar_2013a}.

The $\mK$ matrix defined in \eref{eq: K} is symmetric but the row sum $\mK\vone$ is not 1. Thus, $\mK$ cannot be used directly as a denoising filter because it amplifies or attenuates signals. To normalize the matrix while preserving the symmetry, one can apply the Sinkhorn-Knopp balancing algorithm \cite{Sinkhorn_Knopp_1967, Milanfar_2013b, Chan_Zickler_Lu_2017} to iteratively normalizes the rows and columns of $\mK$ until convergence. When $\mK$ is symmetrized, the resulting matrix is called a symmetric smoothing filter, given by
\begin{equation}
\mW = \mD^{-1/2}\mK\mD^{-1/2},
\label{eq: W symmetric}
\end{equation}
where the diagonal matrix $\mD$ is defined such that $\mW\vone = \vone$ and $\mW^T\vone =  \vone$. Since the columns and rows of $\mW$ sum to 1, $\mW$ is a \emph{symmetric doubly stochastic matrix}. It has been observed that the symmetrized $\mW$ in \eref{eq: W symmetric} has a better denoising performance than its non-symmetric counter part $\mW = \mDtilde^{-1}\mK$ where $\mDtilde = \mbox{diag}(\mK\vone)$. This is attributed to the implicit clustering of the pixels during the Sinkhorn-Knopp balancing algorithm \cite{Chan_Zickler_Lu_2017}, among other reasons such as reduced degree of freedom so that the drop in variance overrides the gain in bias \cite{Milanfar_2013b}.

\subsection{Properties of Graph Filters}
\noindent\textbf{Properties}. The graph filter $\mW$ has a number of properties. We list a few of them here.
\begin{itemize}
\item $\mW$ can be considered as a weighted adjacency matrix of an undirected graph. The $(i,j)$-th element $[\mW]_{ij}$ is the weight on the edge linking node $i$ and node $j$.
\item $0 \le \lambda_{n}(\mW) \le \ldots \le \lambda_{1}(\mW) = 1$. This follows from the fact that $\mW$ is doubly stochastic, and so the eigenvalues are all non-negative.
\item $\mI - \mW$ is the graph Laplacian, with the zero eigenvalue associated to the vector $\frac{1}{\sqrt{n}}\vone$.
\item The regularization defined through the graph Laplacian $\vx^T(\mI - \mW)\vx$ can be interpreted as
\begin{equation}
\vx^T(\mI - \mW)\vx = \sum_{i=1}^n \sum_{j\not=i}^n [\mK]_{ij}\left( \frac{x_i}{\sqrt{[\mD]_{ii}}} - \frac{x_j}{\sqrt{[\mD]_{jj}}}\right)^2,
\end{equation}
which is measuring the smoothness of the signal $\vx$ with the weights defined by the graph $\mW$.
\end{itemize}

To analyze the spectral property of $\mW$, we consider the eigen-decomposition $\mW = \mU\mS\mU^T$ where $\mU$ is the eigenvector matrix and $\mS$ is the eigenvalue matrix. If $\mW$ is not invertible, i.e., $\mbox{rank}(\mW) = r < n$, we can further decompose $\mW$ into
\begin{equation}
\mW =
\begin{bmatrix}
\mU_1 & \mU_2
\end{bmatrix}
\begin{bmatrix}
\mS_1 & \\
& \mS_2
\end{bmatrix}
\begin{bmatrix}
\mU_1^T \\ \mU_2^T
\end{bmatrix},
\label{eq: decompose W}
\end{equation}
where $\mS_2 = \vzero$, and $\mS_1 \in \R^{r \times r}$ is a diagonal matrix storing the non-zero eigenvalues of $\mW$. Since $\mS_2 = \vzero$, we can simplify $\mW$ as $\mW = \mU_1\mS_1\mU_1^T$.

\vspace{2ex}
\noindent\textbf{Pseudo-Linearity}. In the actual PnP ADMM implementation, $\mW$ is pseudo-linear rather than linear. That is, the matrix $\mW$ is a function of the input noisy image $\vy$, or more precisely $\vxhat = \mW(\vy)\vy$ instead of \eref{eq: filtering output}. However, as mentioned in \cite{Romano2015}, the dependency of $\mW$ on $\vy$ is typically small if $\mW$ is estimated from a \emph{pre-filtered} version of $\vy$, for example applying a baseline denoising algorithm to $\vy$ and constructing a new weight matrix $\mW$ based on the baseline estimate. In practice, using a fixed $\mW$ is justified in applications such as \cite{Sreehari_Venkatakrishnan_Wohlberg_2016}, where $\mW$ is updated for the first tens of iterations and is kept fixed for the remaining iterations. Following the same line of arguments, we assume that $\mW$ is pre-defined and is independent of $\vy$. We will occasionally look at the \emph{oracle} case where $\mW$ is estimated from the ground truth $\vx$ so that we can quantify the \emph{best possible} performance of the algorithm. But in general we only require that $\mW$ is independent of $\vy$.

\begin{figure}[t]
\centering
\includegraphics[width=0.8\linewidth]{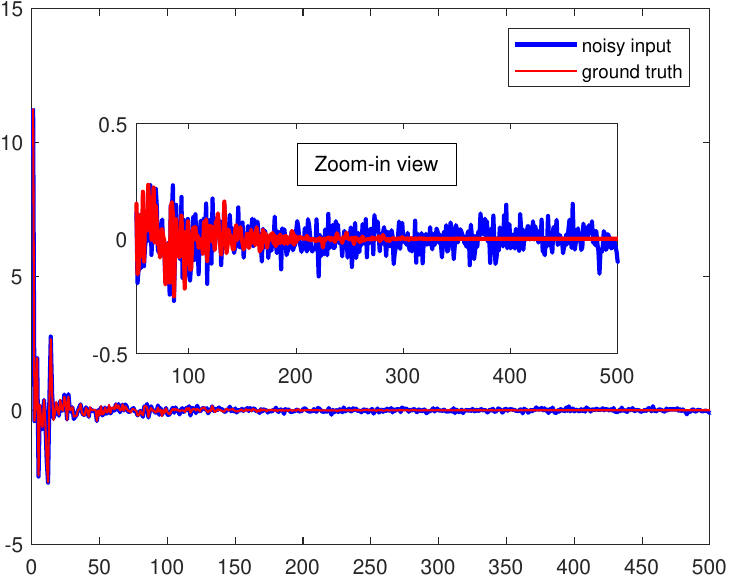}
\caption{Projecting a noisy signal $\vy$ and the ground truth signal $\vx$ by the principle eigenvectors. The two curves show $\mU^T\vy$ and $\mU^T\vx$, where $\mU$ is the eigenvector matrix of $\mW$.}
\label{fig: projection}
\end{figure}

\vspace{2ex}
\noindent\textbf{Energy Concentration}. When applied to a noisy signal $\vy$, $\mW$ projects $\vy$ onto the space spanned by its columns. The effectiveness of this projection is determined by how well $\mW$ is generated. In the oracle setting where $\mW$ is generated from the ground truth signal $\vx$, the first few eigenvectors of $\mW$ is sufficient to capture most of the energy of $\vx$ and leaving the noise in the remaining eigenvectors. This is illustrated in the \fref{fig: projection}, where we consider a 1D signal $\vx$ and its noisy version $\vy$. The weight matrix $\mW$ is constructed from the ground truth $\vx$ using a non-local means kernel defined in \eref{eq: K}. By projecting $\vx$ and $\vy$ using the eigenvector matrix $\mU$, we see that $\mU^T\vx$ has significantly fewer non-zeros than $\mU^T\vy$ for large eigen-indices.

\section{Performance Analysis via MAP}
In this section we present the first set of analytic results. We call it the maximum-a-posteriori (MAP) analysis because we need to explicitly derive the MAP optimization.

\subsection{Existence of $g$}
Given an arbitrary denoiser $\calD_{\sigma}: \R^n \rightarrow \R^n$, the question we ask is whether it is possible to obtain a regularization function $g$ defined in \eref{eq:ADMM2,v}. One way to answer this question is to use a property of the proximal operator by Moreau \cite{Moreau_1965} and show that $\calD_{\sigma}$ corresponds to a proximal map with a convex $g$, a result previously used by Sreehari \emph{et al.} \cite{Sreehari_Venkatakrishnan_Wohlberg_2016} and more recently by Teodoro \emph{et al.} \cite{Teodoro_Bioucas_Figueiredo_2019}.
\begin{lemma}[Moreau \cite{Moreau_1965}]
\label{lemma: existence}
A denoiser $\calD_{\sigma}:\R^n \rightarrow \R^n$ is the proximal map of some function if and only if
\begin{itemize}
\item $\calD_{\sigma}$ is non-expansive, i.e., $\|\calD_{\sigma}(\vx)-\calD_{\sigma}(\vx)\|^2 \le \|\vx-\vy\|^2$ for any $\vx$ and $\vy$, and
\item $\calD_{\sigma}$ is a sub-gradient of some convex function, i.e., there exists $\varphi$ such that $\calD_{\sigma}(\vx) \in \partial \varphi(\vx)$.
\end{itemize}
\end{lemma}
An immediate implication of Lemma 1 is that if $\calD_{\sigma}$ is continuously differentiable, then the subgradient $\partial \varphi$ reduces to the gradient $\nabla \varphi$. Reehorst and Schniter \cite{Reehorst_Schniter_2018} use a classical result by Kantorovitz \cite{Kantorovitz_2016} (in particular Theorem 4.3.8 and 4.3.10) to argue that for such $\varphi$ to exist, $\calD_{\sigma}$ must have a symmetric Jacobian. Therefore, symmetry is necessary for $\varphi$ to exist. Note, however, that symmetry is not sufficient. If $\calD_{\sigma}$ does not have a symmetric Jacobian, then $\varphi$ certainly does not exist. But if $\calD_{\sigma}$ has a symmetric Jacobian, one cannot conclude that $\varphi$ must exist.

Under the context of symmetric smoothing filters, i.e., $\calD_{\sigma}(\vx) = \mW\vx$, both conditions of the Moreau Lemma are satisfied: Non-expansiveness is given by the fact that $\lambda_{1}(\mW) \le 1$, and the sub-gradient condition holds because $\calD_{\sigma}(\vx) = \mW\vx = \nabla\varphi(\vx)$ where $\varphi(\vx) = \frac{1}{2}\vx^T\mW\vx$.

\subsection{Deriving $g$}
We can now derive an explicit form of $g$ when $\calD_{\sigma}(\vx) = \mW\vx$. We will start with a special case where $\mW^{-1}$ exist. The first appearance of this result was reported by Chan \cite{Chan_2016}, and recently generalized by Teodoro \emph{et al.} \cite{Teodoro_Bioucas_Figueiredo_2019}. The proof below is a constructive one, in the sense that we will derive $g$ from $\calD_{\sigma}$ rather than verifying a given expression of $g$.
\begin{theorem}
Let $\mW \in \R^{n \times n}$ be a graph filter, and assume that $\mW^{-1}$ exists. Define $g: \R^n \rightarrow \R$ as
\begin{equation}
g(\vx) = \frac{1}{2 \sigma^2} \vx^T (\mW^{-1} - \mI)\vx.
\label{eq: g expression}
\end{equation}
Then the solution of the minimization of \eref{eq:ADMM2,v} is
\begin{equation}
\vxhat  \;\bydef\;  \argmin{\vx \in \R^n} \;\; g(\vx) + \frac{1}{2\sigma^2}\|\vx - \vxtilde\|^2 = \mW\vxtilde.
\label{eq: nlm v sub}
\end{equation}
\end{theorem}
\begin{proof}
The existence of $g$ is guaranteed by the Moreau Lemma. Taking the gradient of the objective function of \eref{eq: nlm v sub} and equating to zero yields $\nabla g(\vx) + \frac{1}{\sigma^2 }(\vx - \vxtilde) = \vzero$, which is equivalent to
\begin{equation}
(\mI + \sigma^2 \nabla g)(\vx) = \vxtilde.
\label{eq: resolvent 3}
\end{equation}
Define $\mG \in \R^{n \times n}$ as the matrix representation of $\nabla g$ such that $\mG\vx = \nabla g(\vx)$. Then \eref{eq: resolvent 3} can be simplified as $(\mI + \sigma^2 \mG)\vx = \vxtilde$. Substituting $\vx = \mW\vxtilde$, we have that $(\mI + \sigma^2 \mG)\mW\vxtilde = \vxtilde.$ Since this result holds for any $\vxtilde$, we have $(\mI + \sigma^2 \mG)\mW = \mI$, which implies that $\mG = \frac{1}{\sigma^2}(\mW^{-1}-\mI)$. Therefore,
\begin{equation}
\nabla g(\vx) = \mG\vx = \frac{1}{\sigma^2}(\mW^{-1} - \mI)\vx.
\end{equation}
Integrating both sides yields $g(\vx) = \frac{1}{2} \vx^T \mG\vx$.
\end{proof}


If $\mW$ is not invertible, the function $g(\vx)$ should be
\begin{equation}
g(\vx) = \frac{1}{2\sigma^2}\vx^T\mU_1(\mS_1^{-1}-\mI)\mU_1^T\vx + \iota_{\Omega}(\vx),
\label{eq: g with constraint}
\end{equation}
where $\iota_{\Omega}$ denotes an indicator function such that $\iota_{\Omega}(\vx) = 0$ if $\vx \in \Omega$ and $\iota_{\Omega}(\vx) = +\infty$ if $\vx \not\in \Omega$, and the set $\Omega$ is defined as $\Omega = \{ \vx \;|\; \mU_2^T\vx = \vzero\}$. An intuitive argument is that $\vx^T(\mW^{-1}-\mI)\vx$ can be (symbolically) written as
\begin{equation}
\begin{split}
\vx^T(\mW^{-1}-\mI)\vx &= \vx^T\mU_1(\mS_1^{-1}-\mI)\mU_1^T\vx + \\
&\quad\quad + \vx^T\mU_2(\mS_2^{-1}-\mI)\mU_2^T\vx.
\end{split}
\end{equation}
If we put $\mS_2^{-1} = \infty$, then minimizing $\vx^T(\mW^{-1}-\mI)\vx$ will force $\mU_2^T\vx = \vzero$ if we do not allow the trivial solution $\vx = \vzero$. In this case, the function $g$ becomes the one shown in \eref{eq: g with constraint}.

To formally prove the result, we substitute \eref{eq: g with constraint} into \eref{eq: nlm v sub} and check if the optimality condition in \eref{eq: resolvent 3} is satisfied, i.e.,
\begin{equation}
(\mI + \sigma^2 \mG)\mW\vx = \vx,
\label{eq: non invertible condition}
\end{equation}
for any $\vx$. Clearly, if $\vx \not\in \Omega$, then $g(\vx)$ in \eref{eq: g with constraint} is unbounded and so \eref{eq: nlm v sub} does not have a solution. However, if $\vx \in \Omega$, then we can write $\mW = \mU_1\mS_1\mU_1^T$ and substitute $\mG = \frac{1}{\sigma^2}\mU_1(\mS_1^{-1}-\mI)\mU_1^T$ into \eref{eq: non invertible condition}. This will give us
\begin{equation*}
\begin{split}
(\mI + \sigma^2 \mG)\mW\vx
&= (\mI + \mU_1(\mS_1^{-1}-\mI)\mU_1^T)(\mU_1\mS_1\mU_1^T)\vx \\
&= \mU_1\mS_1\mU_1^T\vx + \mU_1(\mI-\mS_1)\mU_1^T\vx \\
&= \vx.
\end{split}
\end{equation*}
Therefore, if $\vx \in \Omega$, the optimality condition in \eref{eq: resolvent 3} is satisfied. The non-invertible case has interesting implications which we shall discuss further in Section~III.F.

\subsection{The Role of $\rho$}
One property we like to investigate is the parameter $\rho$. By substituting \eref{eq: g expression} into \eref{eq:basic optimization 2}, and recalling $\sigma \bydef  \sqrt{\lambda/\rho}$, we can show that PnP ADMM is solving
\begin{equation}
\minimize{\vx} \;\; f(\vx) + \frac{\rho}{2}\vx^T(\mW^{-1}-\mI)\vx.
\label{eq: solution PnP}
\end{equation}

In the literature of graph signal processing, the prior in \eref{eq: solution PnP} is reminiscent to the classical graph Laplacian. An image restoration using graph Laplacian is
\begin{equation}
\minimize{\vx} \;\; f(\vx) + \frac{\lambda}{2}\vx^T(\mI - \mW)\vx.
\label{eq: solution Lps}
\end{equation}
An interesting observation here is that in \eref{eq: solution PnP}, the original regularization parameter $\lambda$ is eliminated because $g$ contains its reciprocal. Thus, $\lambda$ is replaced by the ADMM internal parameter $\rho$. This internal parameter $\rho$ has no influence to the final solution in the classical ADMM because ADMM converges for all $\rho$ when $f$ and $g$ are closed, proper and convex. However, in PnP ADMM, the optimization cost is characterized by $\rho$. Thus, the solution changes with $\rho$, a phenomenon reported in earlier papers, e.g., \cite{Chan_Wang_Elgendy_2017}.

\begin{figure}[h]
\centering
\includegraphics[width=0.9\linewidth]{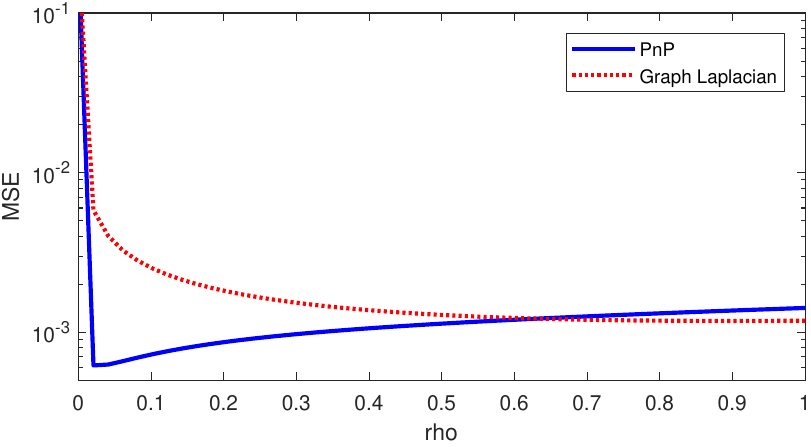}
\caption{Mean squared error of the solutions of \eref{eq: solution PnP} and \eref{eq: solution Lps} as $\rho$ (or $\lambda$) increases. PnP has a lower MSE than graph Laplacian for the best $\rho$ and $\lambda$.}
\vspace{-2ex}
\label{fig: rho}
\end{figure}

We can also ask: For the best $\rho$ and $\lambda$, how would the solutions of \eref{eq: solution PnP} and \eref{eq: solution Lps} behave? \fref{fig: rho} shows an experimental result where we chose the forward model to be $f(\vx) = \frac{1}{2}\|\mA\vx - \vy\|^2$, where the matrix $\mA$ is a convolution matrix constructed from a 3-tap filter $[1, 6, 1]/8$. The specific choice of this 3-tap filter is unimportant. (In MATLAB this is \texttt{A = toeplitz([1 6 1 zeros(1,n-3)])/8}. ) The noisy signal $\vy$ is one column of the \texttt{cameraman} image, corrupted by i.i.d. Gaussian noise with $\sigma_{\eta} = 0.05$. (We assume $\vx \in [0,1]$.) The graph filter $\mW$ is defined by using a non-local mean kernel matrix $\mK$ with $[\mK]_{ij} = \exp\{(i-j)^2/(2h_s^2)\}\exp\{-\|\vx_i-\vx_j\|^2/(2h_r^2)\}$, which is an oracle case. The patch size is $d = 5$. The parameters are set as $h_s = 5$ and $h_r = 0.1$. Sinkhorn-Knopp symmetrization is applied to make sure $\mW$ is symmetric doubly stochastic.

We vary $\rho$ (and equivalently $\lambda$) to evaluate the mean squared error (MSE) for a fixed oracle $\mW$. As shown in \fref{fig: rho}, PnP ADMM and graph Laplacian have different operating regimes for $\rho$ and $\lambda$. PnP ADMM prefers smaller $\rho$, whereas graph Laplacian prefers larger $\lambda$. We will explain this observation in Section 4 using the equilibrium analysis.

\subsection{MSE Analysis for Linear Problems}
To quantify the performance of PnP, we consider a typical linear inverse problem with a data fidelity term
\begin{equation}
f(\vx) = \frac{1}{2}\|\mA\vx - \vy\|^2.
\end{equation}
Ideally, we like to analyze the performance of PnP for an arbitrary $\mA$. However, if $\mA$ is completely arbitrary, the MSE analysis below will involve evaluating eigenvalues of a sum of two Hermitian matrices. There are some classical results on bounding the largest and smallest eigenvalues, e.g.,
\cite{Horn_Johnson_2012}. However, finding the exact eigenvalues of a sum of two Hermitian matrices is an open mathematical problem and currently there is no simple solution \cite{Knutson_Tao_2001}.

In order to ensure analytic tractability, we assume that $\mA$ shares the same eigenvector with $\mW$. That is, $\mA$ has an eigen-decomposition $\mA = \mU\mLambda\mU^T$, where $\mU$ is also the eigenvector of $\mW$. This is summarized in the following assumption.
\begin{assumption}
\label{assumption: A}
We assume that $\mA$ and $\mW$ share the same eigenvector, i.e.,
\begin{equation}
\mA = \mU\mLambda\mU^T, \quad\mbox{and}\quad \mW = \mU\mS\mU^T,
\end{equation}
where $\mLambda = \mbox{diag}(\alpha_1,\ldots,\alpha_n)$ and $\mS = \mbox{diag}(s_1,\ldots,s_n)$, with $0 \le \alpha_i \le 1$ and $0 \le s_i \le 1$ for $i = 1,\ldots,n$.
\end{assumption}
While Assumption 1 is strong, it is reasonable to some extent. If we let $\mU$ be the Fourier transform matrix, then $\mA$ is a convolution matrix representing a blur operation. The corresponding $\mW$ is a lowpass filter. If $\mU$ is the eigenvector of a graph, or graph Fourier transform \cite{Shuman_Narang_Frossard_2013}, then $\mA$ can be chosen as a filtering operation in the graph domain, e.g., diffusion over the graph. As a special case, if we assume $\mLambda = \mI$, then $\mA$ is simplified to $\mA = \mI$, which is a denoising problem.

The problem we are interested in studying is the comparison between the PnP formulation and the graph Laplacian formulation:
\begin{align}
\vxhat_{L} &= \argmin{\vx} \;\; \frac{1}{2}\|\mA\vx - \vy\|^2 + \frac{\lambda}{2} \vx^T (\mI - \mW) \vx, \label{eq: Lps analysis 1}\\
\vxhat_{P} &= \argmin{\vx} \;\; \frac{1}{2}\|\mA\vx - \vy\|^2 + \frac{\rho}{2} \vx^T (\mW^{-1} - \mI) \vx.  \label{eq: PnP analysis 1}
\end{align}
Here, $\mW^{-1}$ is only a short-hand notation of the $g$ defined in \eref{eq: g with constraint}. In the MSE analysis below we will set the eigenvalues to zero if $\mW$ is not invertible.

The solutions of the above two problems are respectively
\begin{align}
\vxhat_{L} &= \left[(\mA^T\mA+\lambda\mI) - \lambda\mW \right]^{-1} \mA^T \vy \bydef \mQ_L^{-1}\mA^T\vy, \label{eq: Lps analysis 2}\\
\vxhat_{P} &= \left[(\mA^T\mA-\rho\mI) + \rho\mW^{-1} \right]^{-1} \mA^T \vy \bydef \mQ_P^{-1}\mA^T\vy, \label{eq: PnP analysis 2}
\end{align}
where the matrices
\begin{align*}
\mQ_L &\bydef (\mA^T\mA+\lambda\mI) - \lambda\mW\\
\mQ_P &\bydef (\mA^T\mA-\rho\mI) + \rho\mW^{-1}
\end{align*}
are used to simplify notations. Letting $\vb \bydef \mU^T\vx$ be the projection of $\vx$ by $\mU$, the MSE of $\vxhat_{L}$ is
{\begin{align}
&\mbox{MSE}_{L}
= \E\|\vxhat_{L}-\vx\|^2 = \E\|\mQ_L^{-1}\mA^T(\mA\vx+\veta)-\vx\|^2 \notag \\
&= \|(\mQ_L^{-1}\mA^T\mA-\mI)\vx\|^2 + \sigma_\eta^2 \mbox{Tr} \left[  \mA \left(\mQ_{L} \mQ_{L}^T\right)^{-1}\mA^T  \right] \notag \\
&= \sum_{i=1}^n \left(\frac{\alpha_i^2}{\alpha_i^2 + \lambda - \lambda s_i}-1\right)^2b_i^2 + \sigma_{\eta}^2\left(\frac{\alpha_i}{\alpha_i^2+\lambda-\lambda s_i}\right)^2\notag \\
&= \sum_{i=1}^n \frac{(\lambda-\lambda s_i)^2b_i^2 + \sigma_{\eta}^2 \alpha_i^2}{(\alpha_i^2 + \lambda - \lambda s_i)^2},
\end{align}}
and the MSE of $\vxhat_{P}$ is
{\begin{align}
&\mbox{MSE}_{P}
= \E\|\vxhat_{P}-\vx\|^2 = \E\|\mQ_P^{-1}\mA^T(\mA\vx+\veta)-\vx\|^2\notag \\
&= \|(\mQ_P^{-1}\mA^T\mA-\mI)\vx\|^2 + \sigma_\eta^2 \mbox{Tr} \left[  \mA \left(\mQ_{P} \mQ_{P}^{T}\right)^{-1}\mA^T  \right]\notag \\
&=  \sum_{i=1}^n \left( \frac{\alpha_i^2 s_i}{\alpha_i^2 s_i - \rho s_i + \rho} - 1\right)^2b_i^2 + \sigma_{\eta}^2\left(\frac{\alpha_i s_i}{\alpha_i^2s_i  - \rho s_i + \rho}\right)^2\notag \\
&= \sum_{i=1}^n \frac{(\rho-\rho s_i)^2b_i^2 + \sigma_{\eta}^2 \alpha_i^2 s_i^2}{(\alpha_i^2 s_i - \rho s_i + \rho)^2}.
\end{align}}

\vspace{-2ex}
Given $\mA$, $\mW$, $\vx$, $\rho$ and $\lambda$, in principle we can completely determine $\mbox{MSE}_{L}$ and $\mbox{MSE}_{P}$. However, instead of resorting to numerical approaches, we analyze the MSEs by looking at the individual terms in the summations above. To this end, we define the $i$-th term as
{\begin{align}
\mbox{MSE}_{L}^i &\bydef \frac{(\lambda -\lambda s_i)^2b_i^2 + \sigma_{\eta}^2 \alpha_i^2}{(\alpha_i^2 + \lambda - \lambda s_i)^2},\\
\mbox{MSE}_{P}^i &\bydef \frac{(\rho-\rho s_i)^2b_i^2 + \sigma_{\eta}^2 \alpha_i^2 s_i^2}{(\alpha_i^2 s_i - \rho s_i + \rho)^2}.
\end{align}}
When $\mW$ is ill-conditioned so that $s_i \rightarrow 0$, we can show that
\begin{align}
\lim_{s_i \rightarrow 0} \mbox{MSE}_{L}^i &= \frac{\lambda^2b_i^2 + \sigma_{\eta}^2 \alpha_i^2}{(\alpha_i^2 + \lambda)^2}, \label{eq:MSE L}\\
\lim_{s_i \rightarrow 0} \mbox{MSE}_{P}^i &= b_i^2. \label{eq:MSE P}
\end{align}
If, in addition to $s_i \rightarrow 0$, we also assume that $b_i \rightarrow 0$ (which typically holds according to the energy concentration of $\mW$ demonstrated in \fref{fig: projection}), then the MSEs become
\begin{align*}
\lim_{b_i, s_i \rightarrow 0} \mbox{MSE}_{L}^i = \frac{\sigma_\eta^2 \alpha_i^2}{(\lambda+\alpha_i^2)^2}, \quad\mbox{and}\quad
\lim_{b_i, s_i \rightarrow 0} \mbox{MSE}_{P}^i = 0.
\end{align*}
{Therefore, if $\alpha_i \not=  0$ (e.g., when $\mA = \mI$ or a matrix with all eigenvalues being positive), then PnP can push the MSE to zero whereas the graph Laplacian has a residue MSE for large eigen-indices $i$.}

\begin{figure*}[!]
\centering
\includegraphics[width=1\linewidth]{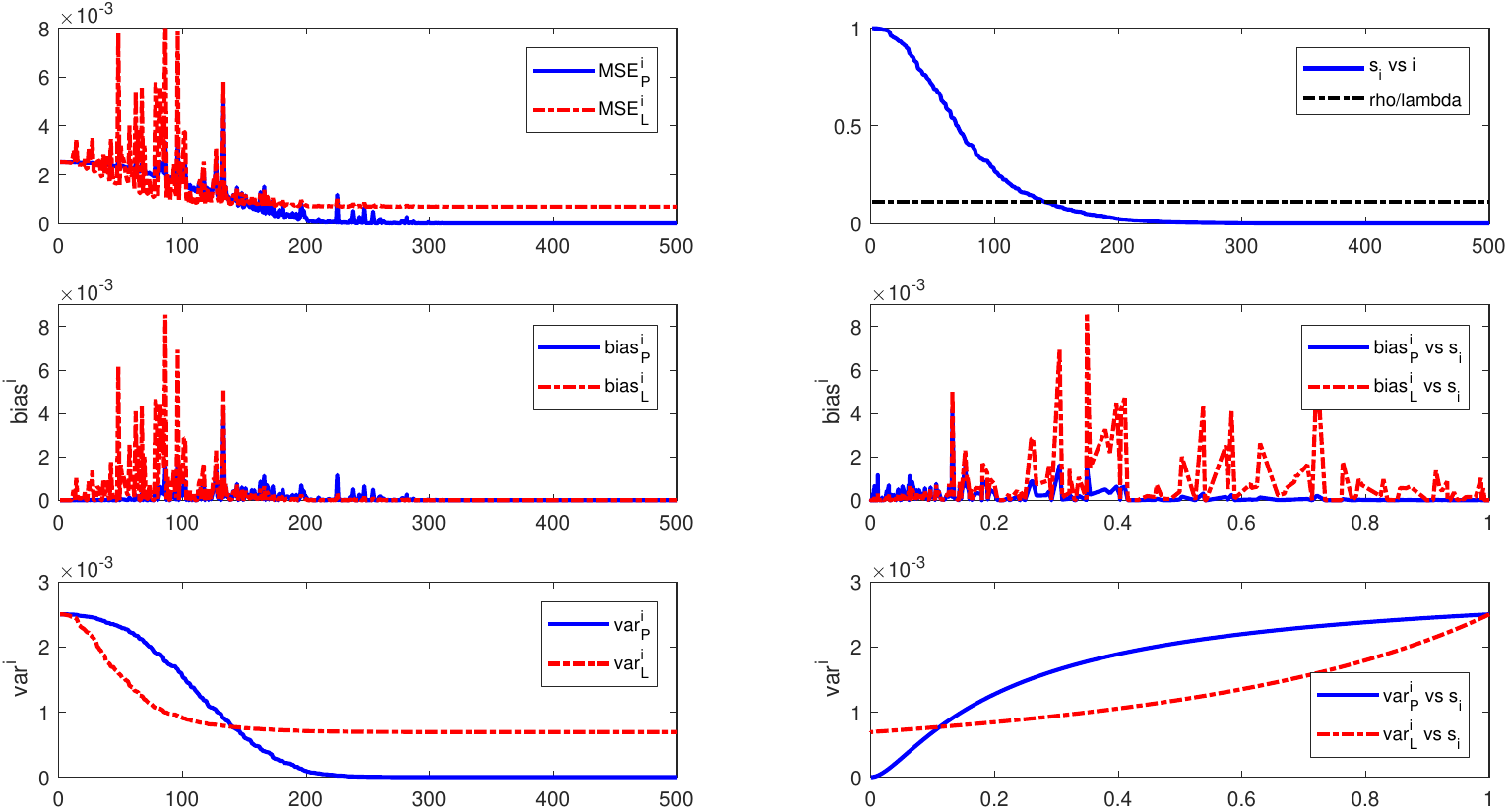}
\caption{Bias and variance of PnP and graph Laplacian. The experiment is based on $\mA = \mI$, $\rho = 0.1$ and $\lambda = 0.9$. The plots on the left show the MSE, bias and variance as a function of the eigen-index $i$. The top right plot shows the eigenvalue $s_i$ as a function of the eigen-index $i$. The two plots on the bottom right show the bias and variance as a function of the eigenvalue $s_i$.}
\vspace{-2ex}
\label{fig: bias var}
\end{figure*}

Our next goal is to analyze the bias and variance so that we can understand the composition of the MSE. A typical breakdown of bias and variance is shown in \fref{fig: bias var}.

The biases of the two methods are
{\begin{equation}
\begin{split}
\mbox{Bias}_L^i &= \frac{(\lambda-\lambda s_i)^2b_i^2}{(\alpha_i^2 + \lambda - \lambda s_i)^2},\\
\mbox{Bias}_P^i &= \frac{(\rho-\rho s_i)^2b_i^2}{(\alpha_i^2 s_i - \rho s_i + \rho)^2}.
\end{split}
\end{equation}}
\begin{proposition}
Assume $\mA = \mU\mLambda\mU^T$, the bias of PnP is greater than the bias of graph Laplacian, i.e.,
\begin{equation}
\mbox{Bias}_L^i \le \mbox{Bias}_P^i
\end{equation}
if and only if the parameters ($\alpha_i,s_i,\lambda,\rho$) satisfy the condition
{\begin{equation}
(\lambda s_i - \rho)\left[s_i\lambda (\alpha_i^2 -2\rho) + \rho(\alpha_i^2+2\lambda)\right] \le 0.
\label{eq:condition for bias}
\end{equation}}
\end{proposition}

\vspace{-2ex}
{\noindent\textbf{Example}. $\mA = \mI$ and $\rho < \frac{1}{2}$. Before we prove the result, it will be useful to look at a special case of \eref{eq:condition for bias}. Consider a denoising problem where $\mA = \mI$ and hence $\alpha_i = 1$ for all $i$. Then, \eref{eq:condition for bias} simplifies to
\begin{equation*}
(\lambda s_i - \rho)[(\lambda-2\lambda\rho)s_i + \rho+2\lambda\rho] \le 0,
\end{equation*}
which holds if the two terms in the product hold different signs. This gives us two cases:
\begin{align*}
\mbox{(i)}  \quad &\lambda s_i - \rho \le 0, \quad\mbox{and}\quad (\lambda-2\lambda\rho)s_i + \rho+2\lambda\rho \ge 0,\\
\mbox{(ii)} \quad &\lambda s_i - \rho \ge 0, \quad\mbox{and}\quad (\lambda-2\lambda\rho)s_i + \rho+2\lambda\rho \le 0.
\end{align*}
Suppose $\rho < \frac{1}{2}$. Then for case (i), we have $1-2\rho > 0$. So $(\lambda-2\lambda\rho)s_i + \rho+2\lambda\rho \ge 0$ is equivalent to $s_i \ge -\frac{\rho}{\lambda}\left(\frac{1+2\lambda}{1-2\rho}\right)$. Since this lower bound is a negative number and $s_i \ge 0$, the inequality is always satisfied. Therefore, $\lambda s_i - \rho \le 0, \;\mbox{and}\; (\lambda-2\lambda\rho)s_i + \rho+2\lambda\rho \ge 0$ can be simplified to $s_i \le \frac{\rho}{\lambda}$.}

{For case (ii), the condition $(\lambda-2\lambda\rho)s_i + \rho+2\lambda\rho \le 0$ is equivalent to $s_i \le -\frac{\rho}{\lambda}\left(\frac{1+2\lambda}{1-2\rho}\right) \le 0$. This is impossible since $s_i \ge 0$. Therefore, case (ii) will never happen when $\rho < \frac{1}{2}$. As a result, if $\mA = \mI$ and $\rho < 1/2$, then $\mbox{Bias}_L^i \le \mbox{Bias}_P^i$ for any $s_i \le \rho/\lambda$.}

\begin{proof}
For notational simplicity we drop the subscript $i$. To show $\mbox{Bias}_L^i - \mbox{Bias}_P^i \le 0$, we cross multiply the numerators and denominators. Since the denominators are positive, it remains to check the combined numerator, which leads us to show
{\begin{equation*}
\begin{split}
&\lambda^2(1-s)^2(\alpha^2s-\rho s + \rho)^2 \\
&\quad\quad - \rho^2(1-s)^2(\alpha^2 + \lambda - \lambda s)^2 \le 0.
\end{split}
\end{equation*}}
Since the left hand side is a difference of two square terms, the expression can be simplified to
{\begin{equation*}
\alpha^2 (1-s)^2[\lambda s - \rho][s\lambda (\alpha^2 -2\rho) + \rho(\alpha^2+2\lambda)] \le 0.
\end{equation*}}
{The terms $\alpha^2$ and $(1-s)^2$ can be dropped because they are both non-negative. Therefore, $\mbox{Bias}_L^i \le \mbox{Bias}_P^i$ if and only if the product of the remaining two terms is non-positive.}
\end{proof}

The analysis of the variance can be carried out in a similar way. The variance of the estimates are
\begin{equation*}
\mbox{Var}_L^i = \frac{\sigma_{\eta}^2 \alpha_i^2}{(\alpha_i^2 + \lambda - \lambda s_i)^2}, \quad\mbox{and}\quad
\mbox{Var}_P^i = \frac{\sigma_{\eta}^2 \alpha_i^2 s_i^2}{(\alpha_i^2 s_i - \rho s_i + \rho)^2}.
\end{equation*}
\begin{proposition}
Assume $\mA = \mU\mLambda\mU^T$, the variance of PnP is less than the variance of graph Laplacian, i.e.,
\begin{equation}
\mbox{Var}_L^i \ge \mbox{Var}_P^i,
\label{eq: variance result}
\end{equation}
if and only if $s_i \le \frac{\rho}{\lambda}$.
\end{proposition}
\begin{proof}
For notational simplicity we drop the subscript $i$. To show $\mbox{Var}_L^i - \mbox{Var}_P^i \ge 0$, we show that
\begin{align*}
&(\alpha^2s - \rho s + \rho)^2 - s^2(\alpha^2 + \lambda - \lambda s)^2 \\
&\quad\quad = \left[2\alpha^2 s + (1-s)(\rho+\lambda s)\right]\left[(1-s)(\rho-\lambda s)\right] \ge 0
\end{align*}
if and only if $\rho-\lambda s \ge 0$, because the rest of the terms are all positive.
\end{proof}
\vspace{-1ex}
Unlike the bias result in \eref{eq:condition for bias} which depends on the spectrum of $\mA$, {the variance result in \eref{eq: variance result} holds for any $\mA = \mU\mLambda\mU^T$}. Now, if we restrict ourselves to $\mA = \mI$, then we can show a \emph{phase transition} point $\rho/\lambda$ such that for any $s_i \le \rho/\lambda$ the variance is lower and for any $s_i \le \rho/\lambda$ the bias is lower. Since $s_i$ is a decreasing function of $i$, it implies that there exists $i^*$ such that for any $i \ge i^*$ we have $s_i \le \rho/\lambda$ and for any $i \le i^*$ we have $s_i \ge \rho/\lambda$. In other words, we can quantify the bias and variance as follows.
\begin{proposition}
Assume $\mA = \mI$ and $\rho < 1/2$. There exists a phase transition point $i^*$ such that
\begin{equation}
\begin{split}
\mbox{Var}_L^i \ge \mbox{Var}_P^i,   &\quad\forall i \ge i^*,\\
\mbox{Bias}_L^i \le \mbox{Bias}_P^i, &\quad\forall i \le i^*.
\end{split}
\end{equation}
The index $i^*$ corresponds to the nearest integer $i$ such that $s_{i} = \rho/\lambda$.
\end{proposition}

To numerically verifying the findings, we conduct the same experiment as in \fref{fig: rho} but with $\mA = \mI$. (Here we set $\rho = 0.1$ and $\lambda = 0.9$ to match a typical value of $\rho$ and $\lambda$.) The results are shown in \fref{fig: bias var}, where we make two sets of plots. On the left hand side, we plot the MSE, the bias, and the variance as a function of the eigen-index $i$. As $i$ increases, there is a clear transition at about $i^* = 140$ such that for any $i \ge i^*$ the variance of PnP is lower, and for any $i \le i^*$ the bias of PnP is lower. On the right hand side of \fref{fig: bias var}, we plot the eigenvalue $s_i$ as a function of $i$ and we overlay the line $\rho/\lambda$ (which is approximately 0.1 in this experiment). The cut off is exactly specified by $i^* = 140$. The bottom right plots of \fref{fig: bias var} shows the bias and variance as a function of $s_i$. If $s_i \le \rho/\lambda$, the bottom plots show that the variance of PnP is lower. If $s_i \ge \rho/\lambda$, the bottom plots show that the bias of PnP is lower.

\vspace{-2ex}
\subsection{Sensitivity of $\mW$}
The performance of PnP is heavily influenced by how well $\mW$ is estimated. If $\mW$ is estimated from the noisy input $\vy$, the performance of PnP will degrade. We conduct an experiment to understand the degradation by constructing a pre-filtered signal via $\vxhat_{\mathrm{pre}} = \vx + \vepsilon$, where $\vx \in \R^n$ is the true signal, and $\vepsilon \sim \calN(\vzero, \sigma_\epsilon^2\mI)$. The graph filter is constructed $\mW$ using $\vxhat_{\mathrm{pre}} $, where the kernel matrix $\mK$ is a non-local mean filter used in \fref{fig: rho}. The constructed $\mW$ is now a function of $\vxhat_{\mathrm{pre}}$. \footnote{An alternative experiment would be to pick a specific baseline denoising algorithm and adjust its internal parameters. However, different baseline denoising algorithms have different characteristics and this could cause bias in the analysis. Thus, we choose to design a pre-filtered signal by adding iid noise.}

\begin{figure}[h]
\centering
\includegraphics[width=0.9\linewidth]{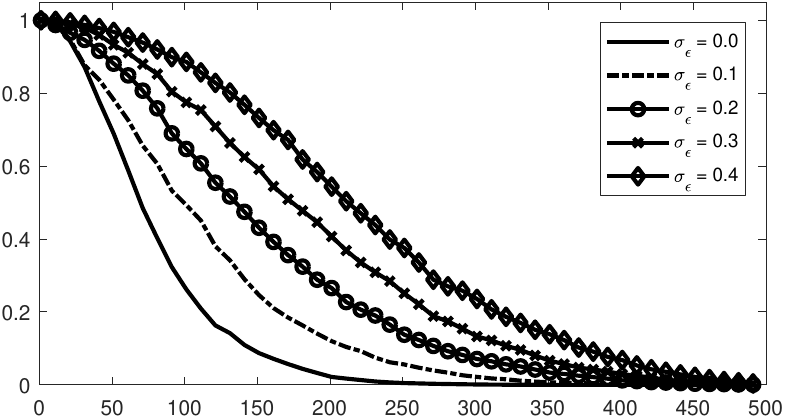}
\caption{Energy concentration of $\mW(\vxhat_{\mathrm{pre}})$ as $\sigma_{\epsilon}$ increases. The plot shows the eigenvalue $s_i$ of $\mW$ as a function of the eigen-index $i$. As $\sigma_{\epsilon}$ increases, the decay rate of $s_i$ reduces and so for the cutoff index $i^*$ shifts towards the right.}
\label{fig: sensitivity}
\end{figure}

When $\mW$ is estimated from $\vxhat_{\mathrm{pre}}$, the energy concentration capability of $\mW$ drops. \fref{fig: sensitivity} shows the eigenvalue $s_i$ as a function $i$ for $\sigma_{\epsilon} \in [0,0.2]$. As we can see from the figure, if we fix $\rho = 0.1$ and $\lambda = 0.9$, the transition point $i^*$ shifts to a large index when noise level $\sigma_{\epsilon}$ increases. Consequently, $\mbox{Var}_L^i \ge \mbox{Var}_P^i$ takes place for a narrower range of $i$'s, whereas $\mbox{Bias}_L^i \le \mbox{Bias}_P^i$ happens for a wider range of $i$'s. Since the variance usually has a bigger margin than the bias when we compare PnP with graph Laplacian (See \fref{fig: bias var}), MSE will become worse as we increase $\sigma_{\epsilon}$.

\begin{figure}[t]
\centering
\includegraphics[width=1\linewidth]{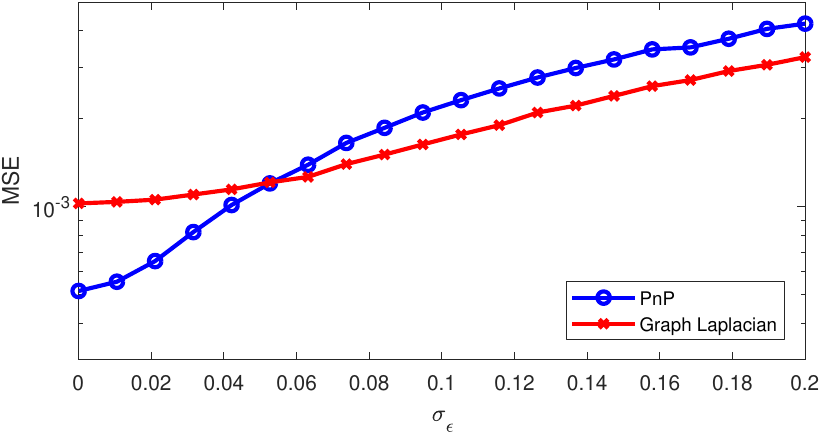}
\caption{Influence of the pre-filtered signal quality to MSE. The experiment uses the $\mA$ and $\mW$ as in \fref{fig: rho}. The parameter $\sigma_\epsilon$ determines the amount of perturbation in $\vxhat_{\mathrm{pre}}$.}
\vspace{-2ex}
\label{fig: prefilter}
\end{figure}

\fref{fig: prefilter} shows another plot of the same experiment, where we compute the MSE by running PnP on a deblurring problem using the $\mA$ and $\mW$ defined in \fref{fig: rho}. We set $\rho = 0.1$ and $\lambda = 0.9$. We observe that PnP has a considerably lower MSE than graph Laplacian when $\sigma_\epsilon$ is small. However, as $\sigma_\epsilon$ increases, $\mbox{MSE}_{P}$ quickly increases and eventually goes above $\mbox{MSE}_{L}$. This shows that while PnP has better performance than graph Laplacian, it is also more sensitive to any error in $\mW$ compared to graph Laplacian.

\subsection{Geometric Interpretation}
\begin{figure}[h]
\centering
\includegraphics[width=0.7\linewidth]{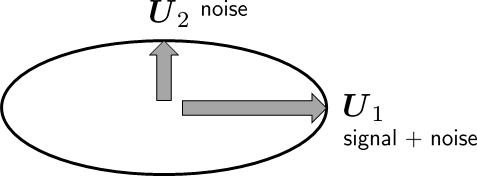}
\caption{PnP ADMM decomposes the optimization variable $\vx$ into two subspaces $\mU_1$ and $\mU_2$. By restricting $\vx$ such that $\mU_2^T\vx = \vzero$, we eliminates the noise living in $\mU_2$. }
\label{fig: eigen diagram}
\end{figure}

A qualitative interpretation of the result can be seen by revisiting $g$ defined in \eref{eq: g with constraint}. For any forward model $f(\vx)$, the optimization solved by PnP is
\begin{align}
\vxhat_{P} &= \argmin{\vx} \;\; f(\vx)+ \frac{\rho}{2} \vx^T \mU_1 (\mS_1^{-1} - \mI)\mU_1^T \vx \notag \\
&\quad\; \mbox{subject to} \;\; \mU_2^T \vx = \vzero.
\label{eq: decompose PnP optimization}
\end{align}
The optimization solved by graph Laplacian is
\begin{align*}
\vxhat_{L} &= \argmin{\vx} \;\; f(\vx)+ \frac{\lambda}{2} \vx^T \mU_1 (\mI - \mS_1)\mU_1^T \vx \\
&\quad\quad\quad\quad\quad\quad   + \frac{\lambda}{2}\vx^T\mU_2(\mI - \mS_2)\mU_2^T\vx,
\end{align*}
which, if we let $\mS_2 = \vzero$, is simplified to
\begin{equation}
\vxhat_L = \argmin{\vx} \;\; f(\vx)+ \frac{\lambda}{2} \left( \vx^T \mU_1 (\mI - \mS_1)\mU_1^T \vx + \|\mU_2^T\vx\|^2\right).
\label{eq: decompose graph optimization}
\end{equation}
The constraint in \eref{eq: decompose PnP optimization} provides a strong restriction on $\vx$. Rather than searching through the entire column space of $\mW$, PnP ADMM only searches for a subspace spanned by eigenvectors $\mU_1$. Because of the energy concentration property of $\mW$, any $\vx$ living in the null space $\mU_2^T\vx = \vzero$ will correspond to noise and no signal. Therefore, by restricting $\mU_2^T\vx = \vzero$ we essentially eliminate the pure noise component observed in $\vy$. This is useful, because it suggests that before we even solve the minimization \eref{eq: decompose PnP optimization}, a large portion of the noise is already cleaned by forcing $\mU_2^T\vx = \vzero$.  \fref{fig: eigen diagram} illustrates the concept pictorially. In graph Laplacian \eref{eq: decompose graph optimization}, the hard constraint $\mU_2^T\vx = \vzero$ becomes a regularization term $\|\mU_2^T\vx\|^2$. However, since $\|\mU_2^T\vx\|^2$ share the same $\lambda$ with $\vx^T \mU_1 (\mI - \mS_1)\mU_1^T \vx$, it can never reach $\|\mU_2^T\vx\|^2 = 0$.

Does a better denoiser give a better PnP ADMM algorithm? The exact answer is unclear because modern denoisers are nonlinear. However, if we focus on linear denoisers, then our analysis suggests that the performance is caused by the rejection of noise before solving the inverse problem. In the extreme case when $\mW$ contains only one non-zero eigenvalue which corresponds to the oracle eigenvector $\vx/\|\vx\|_2$, $\mU_2$ will consist of $n-1$ columns that are orthogonal to $\vx/\|\vx\|_2$. By restricting $\mU_2^T\vx = \vzero$, we are guaranteed to recover $\vx$ exactly. On the other extreme, if $\mW$ is a random matrix or a matrix that does not separate noise from the signal, then the energy concentration property will fail and hence the constraint $\mU_2^T\vx = \vzero$ will have zero or even negative influence.

\section{Performance Analysis via Equilibrium}
The MAP optimization in \eref{eq:basic optimization} is not the only way to obtain $\vxhat$. In \cite{Buzzard_Chan_Bouman_2017}, Buzzard \emph{et al.} generalize PnP ADMM by analyzing the equilibrium state of a dynamical system. In this section we approach the problem from such equilibrium perspective.

\subsection{Consensus Equilibrium}
To understand the concept of consensus equilibrium, we start by defining a pair of functions $F$ and $G$ using \eref{eq:ADMM2,x} and \eref{eq:ADMM2,v} respectively. We can show that at convergence, the PnP ADMM solution $(\vxhat,\vuhat)$ satisfies
\begin{align}
\vxhat &= F(\vxhat + \vuhat), \label{CE, eq1}\\
\vxhat &= G(\vxhat - \vuhat). \label{CE, eq2}
\end{align}
We call such solution $(\vxhat,\vuhat)$ the consensus equilibrium of \eref{CE, eq1}-\eref{CE, eq2}. The following proposition defines the consensus equilibrium of PnP for a linear inverse problem.

\begin{proposition}
\label{prop: CE example}
Consider two operators $F$ and $G$:
\begin{align*}
F(\vx) &= \argmin{\vv\in \R^n} \;\;  \frac{1}{2}\|\mA\vv-\vy\|^2 + \frac{\rho}{2} \|\vv - \vx\|^2,\\
G(\vx) &= \mW\vx
\end{align*}
The consensus equilibrium $(\vxhat,\vuhat)$ satisfies
\begin{subequations}
\begin{align}
&\left(\mW(\mA^T\mA - \rho\mI) + \rho\mI \right)\vxhat = \mW\mA^T\vy, \label{eq: CE example solution a}\\
&\mW\vuhat = (\mW-\mI)\vxhat. \label{eq: CE example solution b}
\end{align}
\end{subequations}
\end{proposition}

\vspace{-2ex}
\begin{proof}
Since $G(\vx) = \mW\vx$, it holds that $G(\vxhat - \vuhat) = \mW(\vxhat - \vuhat)$, which implies that $\mW\vuhat = \mW\vxhat - \vxhat$.
The optimality of $F$ gives
\begin{equation*}
(\mA^T\mA+\rho\mI)\vxhat = \mA^T\vy + \rho(\vxhat + \vuhat).
\end{equation*}
Multiplying both sides by $\mW$ yields
\begin{align*}
\mW(\mA^T\mA+\rho\mI)\vxhat = \mW\mA^T\vy + \rho\mW(\vxhat + \vuhat).
\end{align*}
Substituting $\mW\vuhat = \mW\vxhat - \vxhat$, and rearranging the terms we can show the desired result.
\end{proof}

Proposition~\ref{prop: CE example} is useful in the sense that for a given solution $\vxhat$, we can easily check whether it is a consensus equilibrium point without going through the PnP ADMM algorithm. This allows us to ``create'' alternative optimization formulations and see if the solution is the desired equilibrium point. To illustrate this idea, let us consider two examples.

\vspace{2ex}
\noindent \textbf{Example 1. When $\mA = \mI$.} This is a denoising problem where $f(\vx) = \frac{1}{2}\|\vx - \vy\|^2$. The proposition below shows three optimizations that have the same equilibrium.
\begin{proposition}
\label{prop: equilibrium}
Define the functions $\varphi$, $\psi_1$ and $\psi_2$:
\begin{align}
\varphi(\vx) &= \frac{1}{2}\|\vx - \vy\|^2 + \lambda g(\vx), \label{eq: equilibrium phi}\\
\psi_1(\vx) &= \frac{1}{2}\|\vx - \mW\vy\|^2 - \frac{(1-\rho)}{2}\vx^T (\mI - \mW)\vx, \label{eq: equilibrium psi}\\
\psi_2(\vx) &= \frac{1}{2}\left\| \vx - \left[(1-\rho)\mW + \rho\mI \right]^{-1} \mW \vy\right\|^2,
\end{align}
where $g(\vx)$ is defined in \eref{eq: g with constraint}. Then, the minimizers of these functions all satisfy the equilibrium condition \eref{eq: CE example solution a}-\eref{eq: CE example solution b}.
\end{proposition}

\vspace{-3ex}
\begin{proof}
To prove $\psi_1$, we substitute $\mA = \mI$ into the matrix in \eref{eq: CE example solution a}, and show that $\mW(\mI - \rho\mI) + \rho\mI = \mI - (1-\rho)(\mI-\mW)$. Therefore, the equilibrium condition becomes $\vx = \left[\mI - (1-\rho)(\mI-\mW)\right]^{-1}\mW\vy$, which is equivalent to the first order optimality of \eref{eq: equilibrium psi}. $\psi_2$ is just a restatement of the equilibrium, put in an optimization form.
\end{proof}

In Proposition~\ref{prop: equilibrium}, the first optimization of $\varphi$ is the MAP formulation we saw in Section 3. The third optimization $\psi_2$, is very different from $\varphi$ but it has the same equilibrium as $\varphi$. (Of course, one can argue that $\psi_2$ is useless because it is just a re-statement of the solution.)

The optimization of $\psi_1$ is interesting in the sense that it is different from $\varphi$, but it still resembles the same equilibrium. The objective function $\frac{1}{2}\|\vx - \mW\vy\|^2$ says that instead of trying to minimize the residue between the optimization variable $\vx$ and the noisy observation $\vy$, we minimize $\vx$ with the filtered version $\mW\vy$. For oracle $\mW$'s where pure noise components are isolated at high frequencies, $\mW\vy$ ensures that the pure noise components are eliminated. Therefore, when minimizing $\frac{1}{2}\|\vx - \mW\vy\|^2$, we are guaranteed to only search for the signal component.

The regularization term $-\frac{(1-\rho)}{2}\vx^T (\mI - \mW)\vx$ is a concave function for any $0 \le \rho \le 1$, because $\mI-\mW$ is positive semi-definite. The concave function \emph{injects} high frequency components that are lost by $\mW\vy$. In the extreme cases when $\rho = 1$, we have
\begin{equation*}
\psi_1(\vx) = \frac{1}{2}\|\vx - \mW\vy\|^2,
\end{equation*}
and so the minimizer is just $\vxhat = \mW\vy$. When $\rho = 0$, we have
\begin{align*}
\psi_1(\vx)
&= \frac{1}{2}\|\vx - \mW\vy\|^2 - \frac{1}{2}\vx^T (\mI - \mW)\vx.
\end{align*}
By expanding the terms we can show that
\begin{align*}
\psi_1(\vx)
&= \frac{1}{2}\|\vx\|^2 - \vx^T\mW\vy + \frac{1}{2}\|\mW\vy\|^2 - \frac{1}{2}\|\vx\|^2 + \frac{1}{2}\vx^T\mW\vx\\
&= \frac{1}{2}(\vx - \vy)^T\mW(\vx - \vy) + \frac{1}{2}\vy^T\mW(\mW-\mI)\vy,
\end{align*}
and so the minimizer is $\vxhat = \vy$. In order to control the amount of high frequency injection, a small $\rho$ is more preferred than a large $\rho$. This explains why in \fref{fig: rho} a smaller $\rho$ works better for PnP ADMM.

\vspace{2ex}
\noindent\textbf{Example 2. When $\mA = \mU\mLambda\mU^T$}. In this example, we consider an inverse problem $f(\vx) = \frac{1}{2}\|\mA\vx - \vy\|^2$ for $\mA = \mU\mLambda\mU^T$, where $\mU$ is the eigenvector of $\mW$. The proposition below shows two equivalent formulations.
\begin{proposition}
\label{prop: equilibrium 2}
Let $\mA = \mU\mLambda\mU^T$, where $\mU$ is the eigenvector of $\mW$. Define two functions $\varphi$ and $\psi$
\begin{align}
\varphi(\vx) &= \frac{1}{2}\|\mA\vx - \vy\|^2 + \lambda g(\vx), \label{eq: equilibrium 2 phi}\\
\psi(\vx)    &= \frac{1}{2}(\mA\vx - \vy)^T\mW(\mA\vx-\vy) + \frac{\rho}{2}\vx^T (\mI - \mW)\vx, \label{eq: equilibrium 2 psi}
\end{align}
where $g(\vx)$ is defined in \eref{eq: g with constraint}. Then, the minimizers of these functions all satisfy the equilibrium condition \eref{eq: CE example solution a}-\eref{eq: CE example solution b}.
\end{proposition}

\vspace{-4ex}
\begin{proof}
Since $\mA = \mU\mLambda\mU^T$ and $\mW = \mU\mS\mU^T$, it holds that $\mA$ commutes with $\mW$, i.e., $\mW\mA = \mA\mW$. Thus, the consensus condition can be written as
\begin{align*}
\left[\mA^T\mW \mA + \rho(\mI-\mW)\right]\vxhat = \mA^T\mW\vy,
\end{align*}
which is the first order optimality condition of the optimization
\begin{align*}
\minimize{\vx} \;\; \frac{1}{2}\left\|\mW^{\frac{1}{2}}\mA\vx - \mW^{\frac{1}{2}}\vy\right\|^2 + \frac{\rho}{2}\vx^T(\mI-\mW)\vx.
\end{align*}
Expanding the first term yields the desired result.
\end{proof}

The function $\psi$ can be interpreted as follows. If $\mW$ contains zero eigenvalues, the objective $(\mA\vx-\vy)^T\mW(\mA\vx-\vy)$ will have no cost along those eigenvectors. As a result, since the regularization $\vx^T(\mI-\mW)\vx$ is minimized at $\mU^T\vx = \vzero$, we must have $\mU_2^T\vx = \vzero$ if those eigenvectors with zero eigenvalues.

Rewriting the optimization in terms of $\psi$ also allows us to connect with other works in the literature. In \cite{Kheradmand_Milanfar_2014}, Kheradmand and Milanfar suggest minimizing the following cost when solving a linear inverse problem\footnote{The work in \cite{Kheradmand_Milanfar_2014} considers a general $\mA$ matrix and not only $\mA = \mU\mLambda\mU^T$.}:
\begin{equation}
\begin{split}
\psi(\vx) &= \frac{1}{2}\left\| \mA\vx-\vy \right\|_{\mP}^2+ \frac{\rho}{2} \vx^T(\mI - \mW)\vx,
\end{split}
\end{equation}
where $\mP = \mI + \beta(\mI - \mW)$ for some $\beta \ge -1$. This is more general formulation than ours, because if we let $\beta = -1$, then we obtain \eref{eq: equilibrium 2 psi}.

In summary, the two examples above show that if we are willing to change the forward model $f(\vx)$ by integrating the denoiser into $f(\vx)$, then there are more than optimization formulations which can offer meaningful interpretations while satisfying the equilibrium condition.

\subsection{Performance of Multiple Priors}
Beyond a single denoiser we can extend the equilibrium analysis for multiple denoisers. In this case, the MAP-based optimization is
\begin{equation}
\vxhat = \argmin{\vx \in \R^n} \;\; \varphi(\vx) \bydef \mu_0 f_0(\vx) + \sum_{i=1}^k \mu_i f_i(\vx),
\label{eq: CE main}
\end{equation}
where $f_0: \R^n \rightarrow \R $ is objective function associated with the forward model, and $\{f_i\}_{i=1}^k$ are the regularization functions associated with the prior models (if they exist).

Putting into the equilibrium framework, the operations are
\begin{align*}
F_0(\vx) &= \argmin{\vz \in \R^n} \;\; f_0(\vz) + \frac{\rho}{2}\|\vz - \vx\|^2,\\
F_i(\vx) &= \mW_i \vx, \quad\quad i = 1, \ldots, k.
\end{align*}
At equilibrium, the solution $(\vxhat, \{\vuhat_i\}_{i=0}^k)$ satisfies
\begin{align}
F_i(\vxhat + \vuhat_i) = \vxhat, \quad\mbox{and}\quad \sum_{i=0}^k \mu_i \vuhat_i = \vzero,
\label{eq: CE multi prior condition}
\end{align}
which is a generalization of the case $k=2$ in \eref{CE, eq1} and \eref{CE, eq2}. In \eref{eq: CE multi prior condition}, The mappings $\{F_i\}_{i=0}^k$ are called \emph{agents}, whose relative strengths are controlled by the parameters $\{\mu_i\}_{i=0}^k$. Without loss of generality we assume that $\sum_{i=1}^k \mu_i = 1$, and $\mu_0$ can take any positive value. The intuition behind \eref{eq: CE multi prior condition} is the concept of assembling individual experts to produce an overall better result. For example, $F_i$ could represent image denoisers trained by different image classes, e.g., human faces, buildings and plants, etc. Given an unknown scene, it is likely that one of the denoisers would perform better than the others. Therefore, an ensemble of weak experts can improve the robustness against model mismatch.

To make the following discussion more concrete, let us focus on the forward model $f_0(\vx) = \frac{1}{2}\|\mA\vx - \vy\|^2$ for an arbitrary $\mA$ (not necessarily $\mA = \mU\mLambda\mU^T)$ but assume that $\mW_i$ is invertible. Then, the agent $F_0$ is defined as
\begin{align*}
F_0(\vx) &= \argmin{\vv} \; \frac{1}{2}\|\mA\vv - \vy\|^2 + \frac{\rho}{2}\|\vv - \vx\|^2 \\
&= (\mA^T\mA + \rho\mI)^{-1}(\mA^T\vy + \rho\vx).
\end{align*}
The equilibrium condition can be determined by substituting $\vx = \vxhat + \vuhat_0$, leading to
\begin{equation*}
\vxhat = F_0(\vxhat + \vuhat_0) = (\mA^T\mA + \rho\mI)^{-1}(\mA^T\vy + \rho(\vxhat + \vuhat_0)).
\end{equation*}
Rearranging the terms gives $\vuhat_0 = \frac{1}{\rho}\mA^T\mA\vxhat - \frac{1}{\rho}\mA^T\vy$. For the remaining $k-1$ agents, we have
$$
\vxhat = F_i(\vxhat + \vuhat_i) = \mW_i(\vxhat + \vuhat_i) \;\; \Rightarrow \;\; \vuhat_i = (\mW_i^{-1} - \mI)\vxhat.
$$
Substituting into $\sum_{i=0}^k \mu_i \vuhat_i = \vzero$, we can show that the consensus equilibrium solution satisfies
\begin{equation}
\left(\frac{\mu_0}{\rho}\mA^T\mA + \sum_{i=1}^k \mu_i \left(\mW_i^{-1}-\mI \right) \right)\vxhat = \frac{\mu_0}{\rho}\mA^T\vy,
\label{eq: CE multi prior condition 2}
\end{equation}
for $i = 1,\ldots,k$.

\begin{figure}[h]
\centering
\includegraphics[width=1\linewidth]{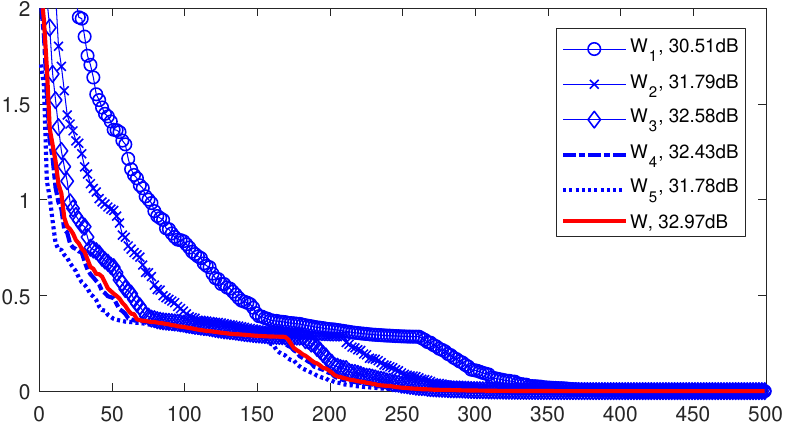}
\caption{Combining multiple priors for a deblurring problem. The curves show the eigenvalues of the inversion matrices defined in \eref{eq: CE inversion matrix 1} and \eref{eq: CE inversion matrix 2}. We set $\rho = 0.005$, $\mu_0 = 1$, $\mu_i = 0.2$ for $i=1,\ldots,5$.}
\label{fig: CE}
\vspace{-2ex}
\end{figure}

To visually see the benefit of the combination, we consider a simple experiment where $\mA$ is the blur matrix used in \fref{fig: rho}. We use $k = 5$ filters $\{\mW_i\}_{i=1}^k$ where the filters $\mW_i$ are generated from kernels with different denoising strengths, see \eref{eq: K}. The combination weights are set as $\mu_i = 1/k$ for simplicity, and we set $\mu_0 = 1$ and $\rho = 0.005$. The equilibrium condition of this combined filter gives a solution of the form
\begin{equation}
\vxhat = \left(\mu_0\mA^T\mA + \rho\left( \sum_{i=1}^k \mu_i \mW_i^{-1} - \mI  \right)\right)^{-1}\mu_0 \mA^T \vy.
\label{eq: CE inversion matrix 1}
\end{equation}
For comparison, we also consider solutions obtained from a single filter. This leads to a solution
\begin{equation}
\vxhat_i = \left(\mu_0\mA^T\mA + \rho( \mW_i^{-1} - \mI)\right)^{-1} \mu_0 \mA^T \vy,
\label{eq: CE inversion matrix 2}
\end{equation}
for $i=1,\ldots,k$. \fref{fig: CE} shows the eigenvalue of the respective inversion matrices. The legend $\mW_i$ in the plot denotes the individual filters we use, and the legend $\mW$ denotes the combined filter. PSNR values of each method is shown in the legend. As one can see, the combined estimate staying in the middle $\mW_3$ and $\mW_4$, which are the two best performing matrices.

\begin{table*}[!]
\footnotesize{
\begin{tabular}{c|cc|cc|cc|cc|cc|cc|cc|cc}
\hline
          & \multicolumn{8}{c|}{Estimated $\mW$} & \multicolumn{8}{c}{Oracle $\mW$} \\
\hline
          & 80\%  &       & 60\%  &       & 40\%  &       & 20\%  &       & 80\%  &       & 60\%  &       & 40\%  &       & 20\%  &     \\
Img       & Lps   & PnP   & Lps   & PnP   & Lps   & PnP   & Lps   & PnP   & Lps   & PnP   & Lps   & PnP   & Lps   & PnP   & Lps   & PnP \\
\hline
Barb      & 26.86 & 28.97 & 26.30 & 27.88 & 25.75 & 26.80 & 23.64 & 24.36 & 50.47 & 53.50 & 48.72 & 52.41 & 47.14 & 50.58 & 44.26 & 49.11\\
Boat      & 26.64 & 28.54 & 25.93 & 27.31 & 25.16 & 25.96 & 23.04 & 23.46 & 50.16 & 51.92 & 48.84 & 51.89 & 47.10 & 50.01 & 43.64 & 48.01\\
Cmra      & 26.39 & 27.88 & 25.30 & 26.33 & 24.13 & 24.81 & 22.01 & 22.26 & 49.67 & 50.53 & 48.32 & 50.20 & 46.99 & 47.92 & 44.28 & 45.62\\
Coup      & 26.55 & 28.46 & 25.87 & 27.25 & 25.19 & 25.98 & 23.22 & 23.67 & 50.25 & 52.72 & 48.18 & 50.81 & 46.98 & 50.79 & 43.96 & 48.51\\
Hous      & 27.28 & 30.52 & 27.22 & 29.83 & 26.95 & 28.67 & 25.07 & 25.82 & 50.66 & 53.56 & 49.16 & 53.39 & 46.81 & 50.72 & 43.63 & 48.52\\
Lena      & 26.98 & 29.77 & 26.62 & 28.70 & 26.25 & 27.57 & 24.28 & 24.94 & 49.62 & 53.13 & 48.29 & 52.18 & 47.45 & 51.44 & 43.93 & 48.15\\
Man       & 26.81 & 29.10 & 26.39 & 28.12 & 25.80 & 26.85 & 23.95 & 24.58 & 50.19 & 53.27 & 49.05 & 52.06 & 47.23 & 51.17 & 44.08 & 49.25\\
Pepp      & 26.36 & 28.41 & 25.64 & 27.10 & 24.67 & 25.51 & 22.66 & 23.21 & 49.69 & 51.12 & 49.20 & 51.26 & 46.61 & 50.64 & 41.98 & 46.77\\
\hline
\end{tabular}}
\caption{Inpainting Experiment. The input images contains of uniformly random sampled pixels at a rate of 80\%, 60\%, 40\%, and 20\%. The reconstruction methods are graph Laplacian (Lps) and PnP ADMM (PnP). The regularization parameter $\rho$ and $\lambda$ are optimized for every image and every configuration. }
\label{tab: inpainting}
\end{table*}

\subsection{Choosing the Weights}
Is it possible to determine a set of weights $\{\mu_i\}_{i=1}^k$ so that they are optimal in some sense? In principle this is doable by minimizing the mean squared error between the $\vxhat$ defined in \eref{eq: CE inversion matrix 1} and the ground truth (if we have access to). However, such approach is computationally prohibited because the $\{\mu_i\}_{i=1}^k$ are located in the denominator of \eref{eq: CE inversion matrix 1}.

A slightly more ad hoc approach is to solve \eref{eq: CE inversion matrix 2} for every $\vxhat_i$, and define $\vxhat$ as a linear combination of them:
\begin{equation}
\vxhat = \sum_{i=1}^k \mu_i \vxhat_i = \sum_{i=1}^k \mu_i \mQ_i^{-1} \mA^T\vy,
\label{eq: CE multi prior condition 4}
\end{equation}
where $\mQ_i \bydef \left(\mA^T\mA + (\rho/\mu_0) ( \mW_i^{-1} - \mI)\right)$. An interpretation of \eref{eq: CE multi prior condition 4} is the solution of an optimization
\begin{align}
&\min_{\vxhat, \{\vxhat_i\}_{i=1}^k} \quad \sum_{i=1}^k \mu_i  \left\| \vxhat - \vxhat_i \right\|^2 \label{eq: CE multi prior condition 5} \\
&\mbox{subject to}  \quad \left(\frac{\mu_0}{\rho}\mA^T\mA + \left(\mW_i^{-1}-\mI \right) \right)\vxhat_i = \frac{\mu_0}{\rho}\mA^T\vy, \notag
\end{align}
which says that among the individual equilibrium solutions $\vxhat_i$, we are finding an average $\vxhat$ that is close to all $\vxhat_i$'s.

The weighted average in \eref{eq: CE multi prior condition 4} provides a simple way of determining $\{\mu_i\}_{i=1}^k$. Assuming that we have access to the ground truth $\vx$, the weights $\{\mu_i\}_{i=1}^k$ can be found by minimizing
\begin{equation}
\min_{\{\mu_i\}_{i=1}^k} \; \left\|\vx - \sum_{i=1}^k \mu_i \vxhat_i \right\|^2 \;\; \mbox{subject to} \;\; \sum_{i=1}^k \mu_i = 1.
\label{eq: CE multi prior condition 6}
\end{equation}
Letting $\vmu = [\mu_1,\ldots,\mu_k]^T$, and defining the matrix $\mSigma \in \R^{k \times k}$ such that $[\mSigma]_{ij} = (\vx-\vxhat_i)^T(\vx-\vxhat_j)$, then \eref{eq: CE multi prior condition 6} can be written as
\begin{equation}
\min_{\vmu} \;\; \vmu^T\mSigma\vmu \;\;\;\; \mbox{subject to} \;\;\;\; \vmu^T\vone = 1.
\label{eq: CE multi prior condition 7}
\end{equation}
Closed-form solution of \eref{eq: CE multi prior condition 7} exists, and is given by
\begin{equation}
\vmu^* = \frac{\mSigma^{-1}\vone}{\vone^T \mSigma^{-1}\vone}.
\end{equation}
Readers interested in the details can refer to \cite{Choi_Elgendy_Chan_2018_TIP} by Choi \emph{et al.}, where they discuss the solution's geometry and its extension to cases where ground truth is not available.

\section{Further Discussions}
We close the paper by discussing a few issues that might be of interest to readers.

\vspace{2ex}
\noindent $\bullet$ \textbf{Experiment on Images}. Thus far the numerical results are all based on 1D signals. Readers who do image processing may wonder how well can the theory be translated to images. To articulate this problem, we consider an inpainting problem which has a forward model as
\begin{equation}
\vy = \mA\vx + \veta.
\end{equation}
In this equation, $\mA \in \R^{m \times n}$ is a sampling matrix which either selects or not selects a pixel from the input image $\vx$. The noise vector $\veta$ is assumed i.i.d. Gaussian with standard deviation $\sigma_{\eta} = 0.05$. Therefore, the objective function is $f(\vx) = \frac{1}{2}\|\mA\vx - \vy\|^2$.

The result of this experiment is shown in Table~\ref{tab: inpainting}, where we compare graph Laplacian with PnP at different sampling ratios (20\%, 40\%, 60\%, 80\%). The $\mW$ matrix is generated from a non-local means kernel using a MATLAB code implementing the Nystrom method in \cite{Talebi_Milanfar_2014}. The prefiltered signal is generated by using the Sheperd interpolation algorithm \cite{Shepard_1968}. We test 8 ``standard'' images and report the PSNR. The regularization parameters $\rho$ (or $\lambda$) are optimized for every image and for every testing condition. Therefore, the PSNRs shown in Table~\ref{tab: inpainting} are the best possible results of the methods. For both oracle and estimated $\mW$, PnP ADMM is consistently better than graph Laplacian, which is coherent to the theoretical arguments.

\vspace{2ex}
\noindent $\bullet$ \textbf{Complexity Consideration}. When we compare \eref{eq: Lps analysis 1} and \eref{eq: Lps analysis 2}, we tend to think that \eref{eq: Lps analysis 1} is computationally less intensive because it is a standard least squares which can be solved using conjugate gradient (or other linear solvers). The PnP solution \eref{eq: Lps analysis 2}, on the other hand, requires spectral decomposition of $\mW$ so that we can obtain the solution via, e.g., by solving an optimization \eref{eq: decompose PnP optimization}. However, a complexity analysis based on this line of argument is not totally correct because in order to obtain the PnP solution we actually use the ADMM algorithm \eref{eq:ADMM2,x}-\eref{eq:ADMM2,u}. Thus, the real complexity bottleneck is at the inversion, i.e., \eref{eq:ADMM2,x}, not the graph filter $\mW$. For specific problems such as deblurring where $f(\vx)$ has structures, e.g., toeplitz matrix, the inversion could be done in $\calO(n \log n)$ via Fourier transform.

\vspace{2ex}
\noindent $\bullet$ \textbf{Statistical Interpretations of $\mW$ and $\mW^{-1}$}. If we regard $\mW$ as some kind of covariance matrix, then the inverse $\mW^{-1}$ is the precision matrix. The $(i,j)$-th element of the covariance matrix measures the \emph{correlation} between node $i$ and node $j$ of the graph, whereas the $(i,j)$-th element of the precision matrix measures the \emph{partial correlation} between node $i$ and node $j$ condition on all the other $n-2$ nodes. Partial correlation is useful in handling situations where a high correlation between $i$ and $j$ is caused by some some third party node $k$. Thus by conditioning on $k$ we can remove the dependency of node $k$, and leaving just the partial correlation between $i$ and $j$.

\begin{figure}[h]
\centering
\begin{tabular}{cc}
\includegraphics[width=0.35\linewidth]{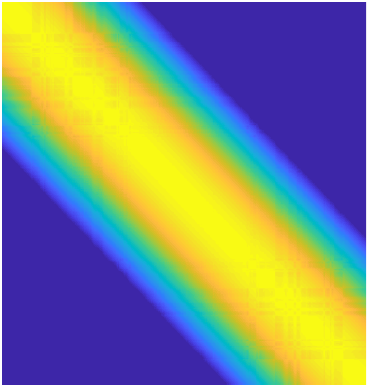}&
\includegraphics[width=0.35\linewidth]{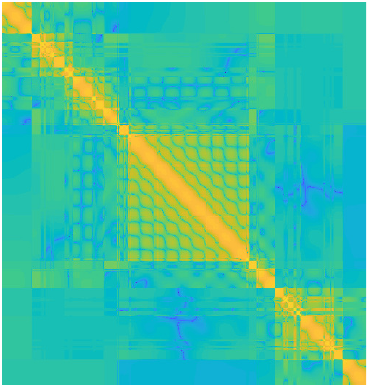}
\end{tabular}
\includegraphics[width=0.8\linewidth]{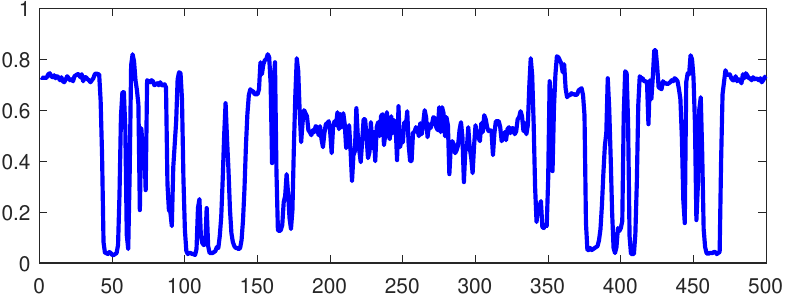}
\caption{[Top Left] $\mW$. [Top Right] $\mW^{+}$, the pseudo-inverse of $\mW$. We plot $\mW$ and $\mW^{+}$ in log absolute scale so that we can visualize them. [Bottom] The ground truth signal $\vx$ used to generate $\mW$.}
\label{fig: W and invW}
\end{figure}

As an example, in \fref{fig: W and invW} we show a ground truth signal $\vx$ used to generate a filter matrix $\mW$. We then compute the pseudo-inverse $\mW^{+}$ by keeping on the 50 leading eigenvectors, and plot its magnitude in the log-scale. (Note: We cannot compute the actual inverse $\mW^{-1}$ when $\mW$ is not invertible.) As we can see, while $\mW$ shows a banded diagonal behavior due to a spatial constraint we put, the inverse matrix $\mW^+$ shows a very strong signal dependent behavior.

\vspace{2ex}
\noindent $\bullet$ \textbf{Connection to RED}. A very similar idea to PnP ADMM is the regularization by denoising (RED) \cite{Romano_Elad_Milanfar_2017}. In RED, the idea is to define the regularization by
\begin{equation}
g(\vx) = \frac{1}{2}\vx^T(\vx - \calD_{\sigma}(\vx)).
\end{equation}
One advantage of such formulation is that we are able to interpret the regularization as the denoiser $\calD_{\sigma}$ is explicitly expressed in $g$. However, as recently discussed by Reehorst and Schniter \cite{Reehorst_Schniter_2018}, while RED offers an interpretable regularization, the existence of $g$ is problematic if $\calD_{\sigma}$ is not symmetric.

If we restrict $\calD_{\sigma}$ to symmetric smoothing filters so that $\calD_{\sigma}(\vx) = \mW\vx$, then the RED regularization is nothing but
\begin{equation*}
g(\vx) = \frac{1}{2}\vx^T(\vx - \mW \vx) = \frac{1}{2}\vx^T (\mI - \mW) \vx,
\end{equation*}
which is exactly the graph Laplacian regularization. As we have shown in the analysis above, under the oracle setting, this does not have the noise rejection capability as PnP has.

\vspace{2ex}
\noindent $\bullet$ \textbf{Dynamics}. In this paper we assume that $\mW$ is defined through some pre-filtered signal, yet in practice $\mW$ is updated with the ADMM iterations. If $\mW$ is updated dynamically, obtaining a closed-form expression of the consensus equilibrium point would become very difficult. This is an open question, and more studies are needed to fully understand how to characterize the solution of this type.

\vspace{2ex}
\noindent $\bullet$ \textbf{Duality}. There is an interesting duality relationship between the graph Laplacian $(\mI-\mW)$ and the PnP prior $(\mW^{-1}-\mI)$. In fact, one can show that
\begin{equation}
(\mI-\mW)^{-1} - (\mW^{-1}-\mI)^{-1} = \mI.
\end{equation}
This result is reminiscent to one of the recent theoretical studies by Feizi et al \cite{Feizi_Suh_Xia_2017} on generative adversarial network (GAN). The analogy is that if $(\mI-\mW)$ represents a discriminator for the real data, then $(\mI-\mW^{-1})$ represents a discriminator for the generated data. It would be interesting to derive similar duality interpretation from a graph perspective.

\vspace{2ex}
\noindent $\bullet$ \textbf{Beyond Image Restoration}. Thus far we have exclusively discussed imaging applications. If we consider $\mW$ as a forward diffusion, then $\mW^{-1}$ is the reversal diffusion. It would be interesting to investigate applications in general graph domain.

\section{Conclusion}
We presented an analysis to understand the performance of the PnP ADMM algorithm. The analysis is based on the class of graph filters. Our findings suggest a few reasons why PnP ADMM performs better than the conventional graph Laplacian regularization {for appropriate regularization parameters}. From the direct mean squared error (MSE) perspective, the zero eigenvalues of the graph filter $\mW$ reject the noise of the observed signal before solving the inverse problem, whereas graph Laplacian retains the noise. As a result, the mean squared error of a PnP estimate has a lower variance than that of the graph Laplacian, and the drop in variance overrides the gain in bias, thus making the overall mean squared error lower.

To further understand the problem, we analyzed the performance from an equilibrium perspective. Instead of deriving the solution from a MAP optimization, we derived the solution by inspecting the equilibrium condition of the ADMM subproblems. We showed that the PnP ADMM solution can be derived from multiple optimization formulations in which all give the same equilibrium solution.

As more sophisticated image restoration modules, e.g., deep neural networks, are being integrated into traditional model based recovery frameworks, we hypothesize that equilibrium will play an important role in analyzing the performance of these methods.

\section{Acknowledgement}
The author thanks Prof. Charles Bouman and Prof. Greg Buzzard of Purdue University for many inspiring discussions, Dr. Brendt Wolhberg of Los Alamos National Lab for offering valuable feedbacks on this paper, and Prof. Gene Cheung of York University for sharing thoughts on graph signal processing. This work is supported, in part, by the National Science Foundation under grants CCF-1763896 and CCF-1718007.

\balance
\bibliographystyle{IEEEbib}
\bibliography{refs}

\end{document}